\documentstyle[prl,aps,epsfig,floats,twocolumn]{revtex} 
\def\br{\vec{r}} 
 
\def\bt{\vec{t}}

\def\bg{\vec{g}} 
\def\bR{\vec{R}}

\def\bnabla{\vec{\nabla}}  
\def\hA{{\hat A}}  
\def\hB{{\hat B}}  
\def\hC{{\hat C}}  
\def\hF{{\hat F}}  
\def\hP{{\bf {\hat P}}}

\def\sig{{\sigma}}  
\def\hsig{{\hat \sigma}}

\def\halfspace{\hskip0.4cm}

\begin{document} 
\draft 
%\large 
 
%@twocolumnfalse\endcsname 
 
\title{Stresses in isostatic granular systems and emergence of force chains} 
\author{Raphael Blumenfeld} 
 
\address{Polymers and Colloids, Cavendish Laboratory, Madingley Road,  
Cambridge CB3 0HE, UK} 
\maketitle 
\date{\today} 
\maketitle 
 
\begin{abstract} 

Progress is reported on several questions that bedevil understanding of granular systems: (i) are the stress equations elliptic, parabolic or hyperbolic? (ii) how can the often-observed force chains be predicted from a first-principles continuous theory? (iii) How to relate insight from isostatic systems to general packings? Explicit equations are derived for the stress components in two dimensions including the dependence on the local structure. The equations are shown to be hyperbolic and their general solutions, as well as the Green function, are found. It is shown that the solutions give rise to force chains and the explicit dependence of the force chains trajectories and magnitudes on the local geometry is predicted.  Direct experimental tests of the predictions are proposed. Finally, a framework is proposed to relate the analysis to non-isostatic and more realistic granular assemblies.

\end{abstract} 
\narrowtext 
 
Granular systems have become a subject of intensive research in recent years both due to their enormous technological importance and the fundamental
theoretical challenges that they pose \cite{GranRev}. In particular, stress transmission has focused much attention following experimental
\cite{ForceChainsExp}\cite{PhotoElastic} and numerical \cite{ForceChainsNum} observations that arching effects give rise to nonuniform stress fields
\cite{Arches} and in particular to chain-like regions of large forces which cannot be straightforwardly described by conventional
approaches\cite{sandpilesmin}.  It has been recognized that to fully understand this phenomenon in general granular packings it is essential to first
understand stress transmission in isostatic systems \cite{Arches}. Isostatic states are configurations of grains in which the intergranular contact
forces can be determined directly from statics, namely, force and torque balance, without reference to stress-strain relations. These states are
characterized by low mean coordination number, which depends on the dimensionality and the roughness of the grains. These states have been shown to be
easy to approach experimentally \cite{BEB}. Several empirical \cite{MEC} and statistical \cite{Arches}\cite{EG} models have been proposed for the
macroscopic equations that govern the stress field in such systems. Very recently, however, the two-dimensional case has been solved from first
principles on the scale of a few grains \cite{BaBl}. The main result of the new theory is an equation that relates directly between the stress tensor
$\hsig$ and a rank-two symmetric fabric tensor $\hP$ which characterizes the microstructure:

\begin{equation}
p_{xx}\sig_{yy} + p_{yy}\sig_{xx} - 2p_{xy}\sig_{xy} = 0 \ .
\label{eq:Ai}
\end{equation}
This 'constitutive' relation is a local manifestation of the torque balance condition beyond the global requirement that $\hsig=\hsig^T$\cite{BaBl}. The tensor $\hP$ can be defined at the grain level as

\begin{equation} p_{ij} = \frac{1}{2}\sum_l \left( r_i^{lg} R_j^{lg} + r_j^{lg} R_i^{lg} \right)
\label{eq:Aii} 
\end{equation} 
where the indices $i, j$ denote the Cartesian components $x, y$, the vectors $\br^{lg}$ and $\bR^{lg}$ are shown in fig. 1, and the sum runs over the loops $l$ that surround grain $g$. Eq. (\ref{eq:Ai}) together with the conventional force and torque balance conditions

\begin{equation}
\bnabla\cdot\hsig = \bg \halfspace ; \halfspace \hsig=\hsig^T
\label{eq:Aiii}
\end{equation}
give a closed set of equations for the stress tensor. Here $\bg = (g_x(\br),g_y(\br))$ is a position-dependent external force field. The only problem with eq. (\ref{eq:Ai}) was that the volume averages of $p_{ij}$ vanish and therefore that it couples between the stress field and {\it fluctuations} in local geometric properties. This made its coarse-graining to macroscopic scales a non-trivial task, but this difficulty was eventually resolved in \cite{Blii}.

\begin{figure} 
\centerline{\psfig{file=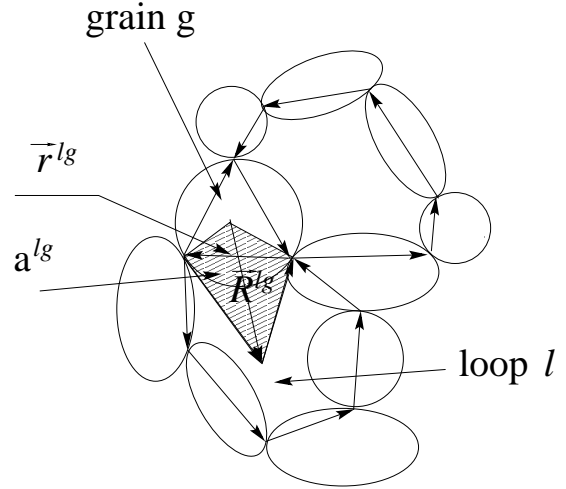,height=6.5cm}} 
\caption{ 
The local geometry around a grain $g$.  The vectors $\br^{lg}$ connect contact points clockwise around each grain $g$ and form loops, e.g. $l$, around physical voids. The vectors $\bR^{lg}$ extend from grain centers to loop centers. The geometry around the grain is characterized by the fabric tensor $\hC^g = \sum_l{\br^{lg}\bR^{lg}}$, whose antisymmetric part gives the area $A^g= \sum_l a^{lg}$ that is the shaded area.  The symmetric part, $\hP$ couples to the stress in eq. (\ref{eq:Ai}).} 
\end{figure} 

Relation (\ref{eq:Ai}) gives a fundamental closure of the stress equations and it is therefore useful to test its implications on two central questions: (a)
Are the stress equations elliptic, parabolic or hyperbolic \cite{ForceChainsExp}\cite{MEC}\cite{MECChaos}\cite{SaChaos}. This question is fundamentally
significant because elasticity theory predicts elliptic equations and any other form implies a significant departure from this paradigm. (b) How can the
continuum theory predict and quantify the emergence of the experimentally observed force chains? Both these issues
are resolved here using the clear geometric interpretation of eq. (\ref{eq:Ai}) from the grain level up.
This Letter is structured as follows: First, explicit equations for the stress components $\sig_{ij}$ are derived and their form is determined. Second, the
general solution of the equations is found and the Green function for an infinite medium is presented. Third, the solution is analysed and it is shown to give
rise to force chains whose individual trajectories can be predicted. Next, experiments are suggested which can directly test the new results. Finally, a short
discussion is presented on the relevance and possible extension of these results to general non-isostatic systems.

To determine the stress field let us consider an isostatic system in mechanical equilibrium, whose microstructure is fully known, and solve
eqs. (\ref{eq:Ai}) and (\ref{eq:Aiii}). To solve for $\sig_{xy}$, for example, substitute $\sig_{yy}$ from eq. (\ref{eq:Ai}) into the balance
eqs. (\ref{eq:Aiii}) then eliminate $\sig_{xx}$ from the two balance conditions by differentiating one with respect to $x$ and the other with respect to $y$. 
Since the coarse-grained $p_{ij}$ are a measure of the {\it fluctuations} of the geometric properties, their gradients are neglected relative to the field gradients. This leaves the equations valid for disordered systems; in ordered periodic structures $p_{ij}=0$ indentically. A similar manipulation for $\sig_{xx}$ and $\sig_{yy}$ gives 

\begin{equation}
\left( p_{xx}\partial_{xx} + 2p_{xy}\partial_{xy} + p_{yy}\partial_{yy} \right) \sig_{ij} = f_{ij}(x,y) 
\label{eq:Aiv} 
\end{equation}
where $\partial_\alpha$ stands for $\partial/\partial \alpha$ and 

\begin{eqnarray}
f_{xy} = & p_{yy}\partial_y g_x + p_{xx}\partial_x g_y \nonumber \\
f_{xx} = & \left(p_{xx}\partial_x + 2p_{xy}\partial_y\right)g_x - p_{xx}\partial_y g_y \nonumber \\
f_{yy} = & \left( p_{yy}\partial_y + 2p_{xy}\partial_x \right) g_y - p_{yy}\partial_x g_x \nonumber
\end{eqnarray}
are functions of the external loading. Note that these functions vanish identically if $\bg$ is constant (e.g. constant gravity). It can be shown that $p_{xy}$
must be finite and, for generality, let us take $p_{xx}$ and $p_{yy}$ to be finite too. When either of these vanishes the solution to eqs. (\ref{eq:Ai}) and
(\ref{eq:Aiii}) simplifies but the qualitative behavior remains unchanged. A key observation is that under the following change of variables

\begin{equation}
\matrix{u \choose v} = {1\  \,\ 0 \choose -\frac{p_{xy}}{S} \, \frac{p_{xx}}{S} } \cdot {x \choose y}
\label{eq:Av}
\end{equation}
where $S = \sqrt{-{\rm det}\hP}$, eq. (\ref{eq:Aiv}) takes the form

\begin{equation}
\left( \partial_{uu} - \partial_{vv} \right) \sig_{ij} = f_{ij} \ .
\label{eq:Avi}
\end{equation}
Eqs. (\ref{eq:Avi}) make it possible to resolve the ongoing dispute regarding the nature of the stress field: They involve only second derivatives and so rule the parabolic form flat out. This suggests that diffusion-like interpretations, which have been proposed to explain the meandering of force chains, are irrelevant. This leaves two possibilities, either the equations are hyperbolic or they are elliptic. From (\ref{eq:Av}) and (\ref{eq:Avi}) we see that the answer hinges on the sign of ${\rm det}\hP$. When it is negative (positive) $v$ is purely real (imaginary) and the equation is hyperbolic (elliptic).
Now note that on the granular level $\hP^g = \sum_l \hP^{lg}$, where $\hP^{lg} = \frac{1}{2} \left(\br^{lg}\bR^{lg} +
\bR^{lg}\br^{lg}\right)$ is the contribution of loop $l$ to $\hP^g$, and the sum runs over the loops around grain $g$ (see fig. 1). It can be readily verified that (-det$\hP^{lg}) = (a^{lg})^2 > 0$ where $a^{lg}$ is the area enclosed by the quadrilateral shown shaded in fig. 1. However, det$\hP^g\neq\sum_l {\rm det}\hP^{lg}$ and therefore it does not enjoy such a convenient interpretation. Rather, a little algebra leads to

\begin{equation}
S = \sqrt{(A^g)^2 - \sum_{ll'} (\br^{lg}\times\br^{l'g})\cdot (\bR^{lg}\times\bR^{l'g}) }
\label{eq:Avii}
\end{equation}
where $A^g = \sum_l a^{lg}$. A careful consideration of the term under the square root leads to the conclusion that its average over a sufficiently large area must be positive definite. For example, in a deformed honeycomb structure with mean separation $c$ between centers of neighboring grains the average of $S$ is $9c^2/4>0$. Nevertheless, it is unclear whether $S$ can fluctuate to negative values on small scales. Observations made on topologically equivalent two-dimensional structures \cite{Bli}, such as liquid crystalline foams \cite{BlCo} and emulsions \cite{JB} reveal no regions with imaginary $v$ down to the cellular (granular) scale. But this may be a consequence of the isotropy of those systems and it is yet unclear whether small clusetrs of grains or foam vertices with $S<0$ can exist in anisotropic systems. That said, the above result on the average of $S$ means that even if such anisotropic systems exist, clusters of 'elliptic defects' cannot survive above a certain lengthscale and therefore eqs. (\ref{eq:Avi}) are indeed {\it hyperbolic on macroscopic scales}.

Turning to the analysis of eqs. (\ref{eq:Avi}), their general solution is

\begin{eqnarray}
\hsig & = \hA^+ \Phi(\eta= v-u) + \hA^- \Psi(\zeta=v+u) + \hB^+\eta + \hB^-\zeta \nonumber \\
& + \frac{1}{4}\int^\eta\int^\zeta \hF(\eta',\zeta') d\eta' d\zeta'
\label{eq:Aix} 
\end{eqnarray}
Here $\hF$ is a rank-two matrix whose components are $f_{ij}$, 

$$\matrix{\hA^{\pm}} = {\frac{p_{xx}}{\alpha^{\pm}} \,1 \choose 1 \, \frac{\alpha^{\pm}}{p_{xx}} } \ \ \ ; \ \ \ \matrix{\hB^{\pm}} =
{\frac{\alpha^{\mp}}{p_{yy}} \,1 \choose 1 \, \ \frac{\alpha^{\pm}}{p_{xx}} } b^{\pm}_{xy}$$
and $\alpha^{\pm} = p_{xy} \pm S$. In solution (\ref{eq:Aix}) $\Phi(\eta)$, $\Psi(\zeta)$ and the coefficients $b^{\pm}_{xy}$ are determined by the boundary data. The lines of constant

\begin{equation}
\eta = [p_{xx}y - \alpha^+x]/S \ \ \ ; \ \ \ 
\zeta = [p_{xx}y - \alpha^-x]/S 
\label{eq:Aixa} 
\end{equation} 
are the characteristic curves of the equations and play a significant role in the behavior of the stress. The Green function of eq. (\ref{eq:Avi}) when $f_{ij}$ is replaced by a $\delta$-function at $(u_0,v_0)$ within an infinite medium is

\begin{eqnarray}
G(u,v;u_0,v_0) & = \frac{1}{2} [ H(v - v_0 - (u - u_0)) \nonumber \\
& + H(v - v_0 + (u - u_0))] 
\label{eq:Ax}
\end{eqnarray}
where $H$ is the Heavyside step function.

For illustration, consider a system occupying the half-plane $x>0$ that is loaded along the $y$-axis with the boundary conditions
$\sig_{ij}(x=0,y)=U_{ij}(y)$ and $\partial_x\sig_{ij}(x=0,y)=V_{ij}(y)$. The choice of these boundary data follows the hyperbolic nature of the
eqs. (6). The system is also presumed to be under a constant field $\bg$. It
is convenient to first convert the boundary conditions to the $u-v$ plane, where they become the Cauchy data. The solution for the stress then becomes

\begin{eqnarray}
\sig_{ij} & = \frac{1}{2}\left[ U_{ij}\left(\frac{S}{p_{xx}}\eta\right) + U_{ij}\left(\frac{S}{p_{xx}} 
\zeta\right) \right] + \nonumber \\
& \frac{1}{2}\int^{\frac{S\zeta}{p_{xx}}}_{\frac{S\eta}{p_{xx}}} \left[ V_{ij}(t) + 
\frac{p_{xy}}{S}U'_{ij}(t)\right] dt 
\label{eq:Axi} 
\end{eqnarray}
where $U'_{ij}$ is the derivative of $U_{ij}$ with respect to its argument. The lack of symmetry in (\ref{eq:Axi}) upon interchanging $x$ and $y$ is a result of the asymmetric choice of the variables $u$ and $v$ with respect to $x$ and $y$. The acid test of this solution is that it can explain the emergence of force chains so often observed in planar systems \cite{ForceChainsExp}-\cite{ForceChainsNum}: Suppose that the boundary loading on the granular system is localized, namely, $V_{ij}$ and $U'_{ij}$ are very narrow for all $\sig_{ij}$.  This is typically what happens when grains, which cannot form a perfect straight line at the boundary, are compressed by a flat surface, which localizes the loading on protruding grains. From
(\ref{eq:Axi}) we see that the localized load propagates into the system along the characteristic curves $\eta = c_\eta$ and $\zeta = c_\zeta$, as shown schematically in fig. 2. In the $x-y$ plane the characteristic curves meander due to the local dependence on the fluctuating geometric tensor $\hP$ and the propagation of the forces into the system will correspondingly deviate from straight trajectories. For a concrete example, suppose that for the $xx$ component the boundary conditions are $U_{xx}(0,y)=e^{-y^2/2d^2}$, with $d$ the size of the loading region, and $V_{xx}(0,y)=0$. Then

\begin{equation}
\sig_{xx} = \frac{1}{2}\left[\left(1 + \frac{p_{xy}}{S}\right) e^{-\frac{S\zeta^2}{2d^2p_{xy}^2}} +  \left(1 - \frac{p_{xy}}{S}\right)e^{-\frac{S\eta^2}{2d^2p_{xx}^2}} \right] 
\label{eq:Axii}
\end{equation}
comprises of two bell-shaped peaks that propagate along the two characteristic curves. It is now possible to calculate the force along the
characteristics: $F_\eta = \hsig\cdot\bt_\eta$ and $F_\zeta = \hsig\cdot\bt_\zeta$, where $\bt_\eta$ and $\bt_\zeta$ are the unit tangents along the
curves. Thresholding and visualizing now the force field magnitude there will emerge two lines tracing the characteristics, whose thicknesses depend on
the threshold level. These are the {\it force chains}! One may argue that force chains can be observed on scales of one or two grains while eqs.
(\ref{eq:Ai}) and (\ref{eq:Aiii}) are continuous and coarse-grained and so is their solution. But the only constitutive data in the solution involves
the fabric tensor $\hP$ and this tensor is well-defined down to the granular scale. Therefore there is no reason that solutions (\ref{eq:Axi}) should
not apply also down to this scale. This analysis not only explains the force chains, it also predicts their individual trajectories as a function of
the local geometry. For example, $\bt_\eta=(p_{xx},\alpha^+)/\sqrt{p_{xx}^2+(\alpha^+)^2}$, $\bt_\zeta=(p_{xx},\alpha^-)/\sqrt{p_{xx}^2+(\alpha^-)^2}$.

More general forms of boundary loading can be regarded as a superposition of many localized forces such as the one described above. These will initiate pairs of chains that will propagate into the material \cite{PhotoElastic}.

When $p_{xx}=0$ or $p_{yy}=0$ the characteristic curves take a slightly different form. For example, for $p_{xx}=0$
$\eta=y-(1+p_{yy})x/(2p_{xy})$, $\zeta=y-(1-p_{yy})x/(2p_{xy})$ and the functions $f_{ij}$ change slightly. Nevertheless, the qualitative behavior of propagation from the boundaries into the system along the characteristic curves remains exactly the same.

The above predictions concerning the force chains can be used to test the theory experimentally as follows: Apply a localized load on the boundary of a granular (or cellular) system where it is possible to trace the force chains. This can be achieved, e.g., using photoelastic grains \cite{PhotoElastic} or liquid-crystalline foams \cite{BlCo}\cite{CourtyEarlyWorks} between crossed polarizers. By scanning the local microstructure one can compute the fabric tensor $\hP$ and then, using expressions (\ref{eq:Aixa}) for $\eta$ and $\zeta$, one can compute the trajectories of the characteristic curves in the vicinity of the concentrated load. These trajectories can be checked for coincidence with the observed trajectories of the force chains to yield a straightforward test of the theory. 

Finally, what is the relevance of these results to general granular packings? The following offers a framework for the extension of the above theory to
all dry granular systems. The ideas presented are based on insight from a recent experiment on inertia-free growing 2D granular piles of rigid and rough
grains \cite{BEB}. Two results of that experiments are of major significance: One is that the isostatic state (IS) can be approached arbitrarily closely.
The other is that this state can be regarded as a critical point where a certain lengthscale diverges. This lengthscale is the size of a yield front
which moves ahead of the consolidating pile wherein grains constantly shift before they finally consolidate. Significantly, this
length characterizes the range of rearrangement as the system is perturbed locally. At the IS the mean coordination number is three, the material is marginally rigid, grains are just stable and a small displacement anywhere causes rearrangement far away. The range of rearrangement is the correlation length, $\xi$. A critical-like behaviour has also been observed in other experiments \cite{Criticality}. As more contacts are made the system moves away from the IS, the density increases, grains are better supported and rearrangement is more confined. Since rearrangement is mediated by forces it follows that the correlation length also describes the typical length
of force chains. Thus, at the IS force chain are expected to span distances comparable to the system size. Indeed, {\it infinitely long} chains is exactly the prediction of eqs. (\ref{eq:Ax}) and (\ref{eq:Axi}). Away from the IS {\it elastic} domains form in which the stress equations are elliptic.  When a force chain is incident on an elastic region it can be shown that it splits into several weaker forces. The idea of the IS as a critical point has already been entertained in the literature \cite{Mo} but within elasticity theory, thus failing to take into consideration the hyperbolic nature of force chains. It is proposed
here that granular systems exhibiting finite force chains are neither at marginal rigidity nor fully elastic, but rather a mixture of these two {\it states}. This means, for example, that $\xi$ would decay as a power of the density difference from the
critical density identified in \cite{BEB}. 
Thus, for a complete theory of stresses in granular media it is essential to address such two-phase materials. Now that we possess a
theory of the isostatic state, it is this author's belief that the extension to two-phase systems is well within reach and a detailed analysis along these lines is under preparation \cite{MRSrelevance}.

To conclude, this paper reported several results: 
1) For isostatic systems it has been found that on large scales all the stress components follow an identical {\it hyperbolic} equation, but with different source terms. 
2) The general solutions for the stress field were found and analysed and the Green function for an infinite medium was presented. 
3) It was shown that the solutions give rise to {\it force chains} whose individual trajectories can be predicted explicitly. 
4) Experiments were suggested which can directly test the validity of the new results. 
5) It is proposed that general granular packings are in fact two-phase mixtures of isostatic and elastic regions and a framework to describe such materials has been suggested.

\begin{figure} 
\centerline{\psfig{file=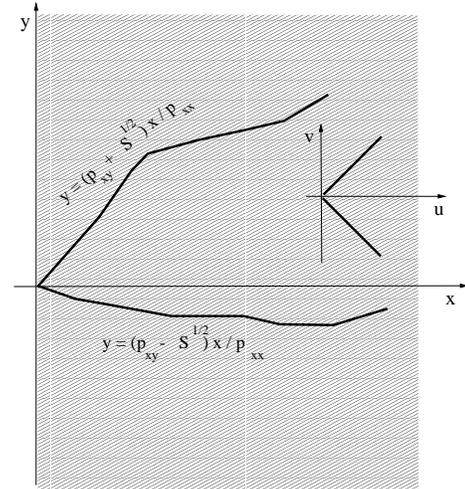,height=6.5cm}} 
\caption{ 
Two force chains bifurcating from a localized force applied to the boundary. The trajectories follow the local characteristic curves $\eta=$, $\zeta=$ constant. The characteristics, which are straight lines in the $u-v$ plane (inset), meander in the $x-y$ plane according to the local values of the components of $\hP$.} 
\end{figure}

\end{document}